\RequirePackage{fix-cm}
\documentclass[smallextended]{svjour3}       
\smartqed  
\usepackage{graphicx}
\usepackage{color}
\begin{document}

\title{Breather wave and lump-type solutions of new (3+1)-dimensional
Boiti-Leon-Manna-Pempinelli equation in incompressible fluid}


\author{ Jian-Guo Liu$^{*}$, Abdul-Majid Wazwaz
}


\institute{Jian-Guo Liu(*Corresponding author) \at
              College of Computer, Jiangxi University of
Traditional Chinese Medicine, Jiangxi 330004, China
Tel.: +8613970042436\\
              Fax: +86079187119019\\
              \email{20101059@jxutcm.edu.cn}\\
\and Abdul-Majid Wazwaz \at Department of Mathematics, Saint Xavier
University, Chicago, IL 60655, USA\\
\email{wazwaz@sxu.edu} }

\date{Received: date / Accepted: date}

\maketitle

\begin{abstract}
Under investigation is a new (3+1)-dimensional
Boiti-Leon-Manna-Pempinelli equation. The main results are listed as
follows:
 (i)  lump solutions;
 (ii) Interaction solutions between lump wave and solitary
waves;
 (iii)  Interaction solutions between lump wave and periodic waves;
 (iv) Breather wave solutions. Furthermore, graphical representation of
 all
 solutions is studied and shown in some 3D- and contour plots.

\keywords{lump solutions, breather wave solutions, solitary waves,
periodic waves.}
\subclass{35C08 \and 68M07 \and 33F10}
\end{abstract}

\section{Introduction}
\label{intro} \quad The (3+1)-dimensional
Boiti-Leon-Manna-Pempinelli (BLMP) equation can be used to describe
the wave propagation in incompressible fluid [1] and the interaction
of the Riemann wave [2]. Many results about BLMP equation have been
obtained. Darvishi [2] presented the stair and step Soliton
Solutions. Zuo [3] derived the bilinear form,  Lax pairs and
B\"{a}cklund transformations. Tang [4] obtained the new
periodic-wave solutions. Liu [5,6] given the new three-wave
solutions and new non-traveling wave solutions. Mabrouk [7] obtained
some new analytical solutions. Li [8] presented the multiple-lump
waves solutions. Osman [9] derived lump waves, breather waves, mixed
waves, and multi-soliton wave solutions. Peng [10] studied the
breather waves and rational solutions. Xu [11] given the
Painlev\'{e} analysis, lump-kink solutions and localized excitation
solutions.

\quad In this paper,  a new (3+1)-dimensional BLMP equation is
studied as follows [12]
\begin{eqnarray}
(u_x+u_y+u_z)_t+(u_x+u_y+u_z)_{xxx}+(u_x(u_x+u_y+u_z))_x=0,
\end{eqnarray}where $u=u(x,y,z,t)$. The new  BLMP equation was first proposed by Wazwaz [12]. The integrability, compatibility
conditions, multiple soliton solutions and multiple complex soliton
solutions were discussed by Painlev\'{e} test and Hirota's direct
method.

\quad Based on the transformation {\begin{eqnarray}
u=-2\,[ln\xi(x,y,z,t)]_x,
\end{eqnarray}
}the new BLMP equation has the bilinear form{\begin{eqnarray}
[D_x^4+D_y D_x^3+D_z D_x^3+D_t\,D_x+D_t\,D_y+D_t\,D_z] \xi\cdot
\xi=0.
\end{eqnarray}
}
 \quad The organization of this paper is as follows. Section 2
gives the lump solutions; Section 3 obtains the interaction
solutions between lump wave and solitary waves; Section 4 derives
the interaction solutions  between lump wave and periodic waves;
Section 5 presents the breather wave solutions. All results  have
been verified to be correct by using Mathematica software [13-25].
Section 6 gives the conclusion.

\section{Lump solutions} \label{sec:2}
\quad Generally speaking, the lump solutions of nonlinear integrable
equations can be assumed as follows{\begin{eqnarray} \xi&=&\gamma
_3+\left(\alpha _4 t+\alpha _1 x+\alpha _2 y+\alpha _3
z\right){}^2+\left(\beta _4 t+\beta _1
   x+\beta _2 y+\beta _3 z\right){}^2,
\end{eqnarray}}where $\alpha_i(1\leq i \leq 4)$, $\beta_i(1\leq i \leq 4)$ and $\gamma_3$ are
undetermined real parameters. Substituting  Eq. (4) into Eq. (3) and
making the coefficients of $x^2$, $y^2$, $xy$, $xz$ et al. be zero,
undetermined real parameters in Eq. (4) have the following  results
\begin{eqnarray}&&(I)\,\,\, \alpha_3=-\alpha _1-\alpha _2, \beta_2= -\beta _1-\beta
_3.\\
&&(II)\,\,\,\beta_1=-\frac{\alpha _1 \alpha _4}{\beta _4},
\alpha_3=\frac{\alpha _1 \beta _3}{\beta _1},
 \beta_2=-\frac{\alpha _2 \beta _3}{\alpha _1+\alpha _2}, \beta_4=\frac{\left(\alpha _1+\alpha _2\right) \alpha _4}{\beta _3}.\\
&&(III)\,\,\,\alpha_3=\frac{\alpha _1 \beta _3}{\beta _1},
\beta_1=-\frac{\alpha _1 \beta _3}{\alpha _1+\alpha _2},
\beta_2=-\frac{\alpha _2 \beta _3}{\alpha _1+\alpha _2}.\\
&&(IV)\,\,\,\alpha_3=-\frac{\beta _1 \beta _3}{\alpha _1},
\beta_3=\frac{\alpha _1^2+\alpha _2 \alpha _1}{\beta _1},
\beta_2=-\frac{\alpha _1 \left(\alpha _1+\alpha _2\right)+\beta _1^2}{\beta _1}.\\
&&(V)\,\,\,\alpha_2=-\alpha _1, \beta_2=-\beta _1, \alpha_3=\beta_3=0.\\
&&(VI)\,\,\,\alpha_4=\beta_4=0, \beta_3=-\frac{\alpha _1
\left(\alpha _1+\alpha _2+\alpha _3\right)+\beta _1 \left(\beta
_1+\beta
   _2\right)}{\beta _1}.\\
&&(VII)\,\,\, \alpha_3=-\alpha _1-\alpha _2, \beta_1= -\beta
_2-\beta
_3, \beta_4=\frac{2 \alpha _1 \alpha _4 \beta _1}{\alpha _1^2-\beta _1^2}.\\
&&(VIII)\,\,\, \alpha_3=-\alpha _1-\alpha _2, \beta_1= \frac{\alpha
_1 \beta _2}{\alpha _2},\nonumber\\ &&\beta_3=-\frac{\left(\alpha
_1+\alpha _2\right) \beta _2}{\alpha _2}, \beta_4=\frac{2 \alpha _1
\alpha _4 \beta _1}{\alpha _1^2-\beta _1^2}.
\end{eqnarray}Substituting Eqs. (5)-(12) into Eq. (2) and Eq. (4),
eight different types of lump solutions are derived. As an example,
substituting Eq. (5) into Eq. (2) and Eq. (4), we have
\begin{eqnarray}u&=&-[2 [2 \alpha _1 [\alpha _4 t+\alpha _1 x+\alpha _2 y-\left(\alpha _1+\alpha
   _2\right) z]+2 \beta _1 [\beta _4 t+\beta _1 x\nonumber\\&-&\left(\beta _1+\beta _3\right) y+\beta
   _3 z]]]/[\gamma _3+[\alpha _4 t+\alpha _1 x+\alpha _2 y-\left(\alpha _1+\alpha
   _2\right) z]{}^2\nonumber\\&+&[\beta _4 t+\beta _1 x-\left(\beta _1+\beta _3\right) y+\beta _3
   z]{}^2].
\end{eqnarray}Graphical representation of lump solution (13) is
shown in Fig. 1. It's obvious that there is a peak and a bottom in
Fig. 1, and they are symmetric. In the peak, solution (13) has a
maximum $2 \sqrt{5}$ at $x= -1/\sqrt{5}$ and $y = 0$. In the bottom,
solution (13) has a minimum $-2 \sqrt{5}$ at $x= 1/\sqrt{5}$ and $y
= 0$.

\includegraphics[scale=0.55,bb=-10 270 10 10]{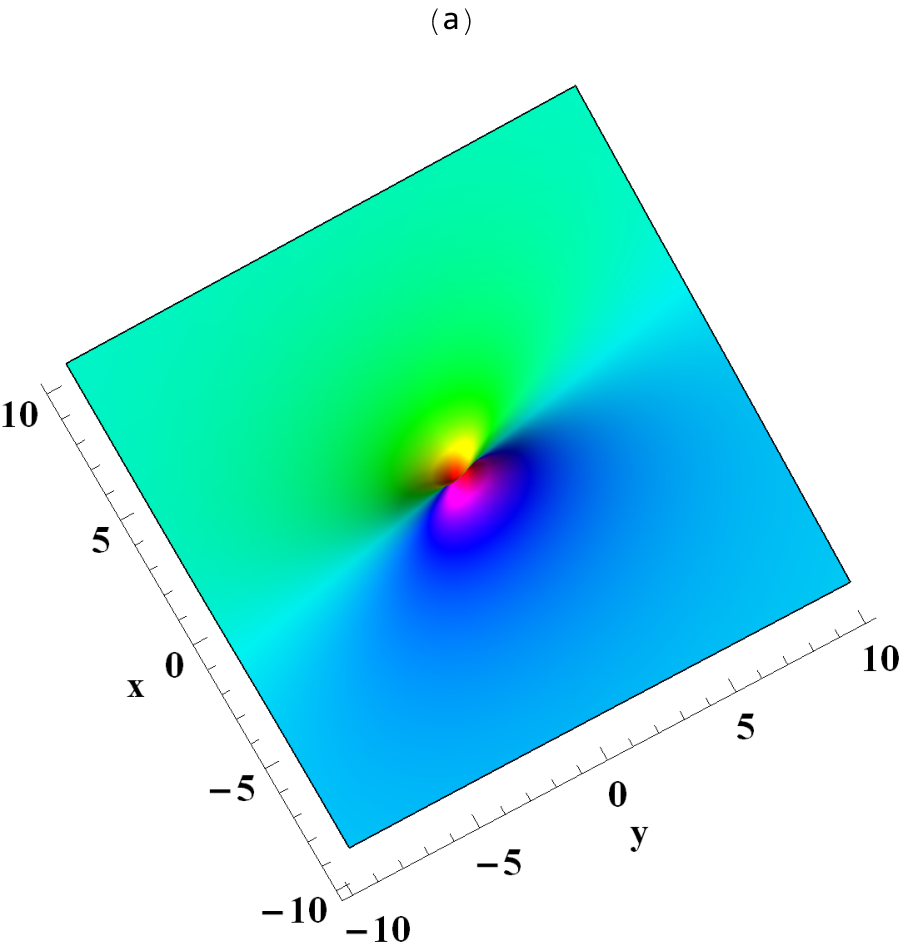}
\includegraphics[scale=0.45,bb=-270 310 10 10]{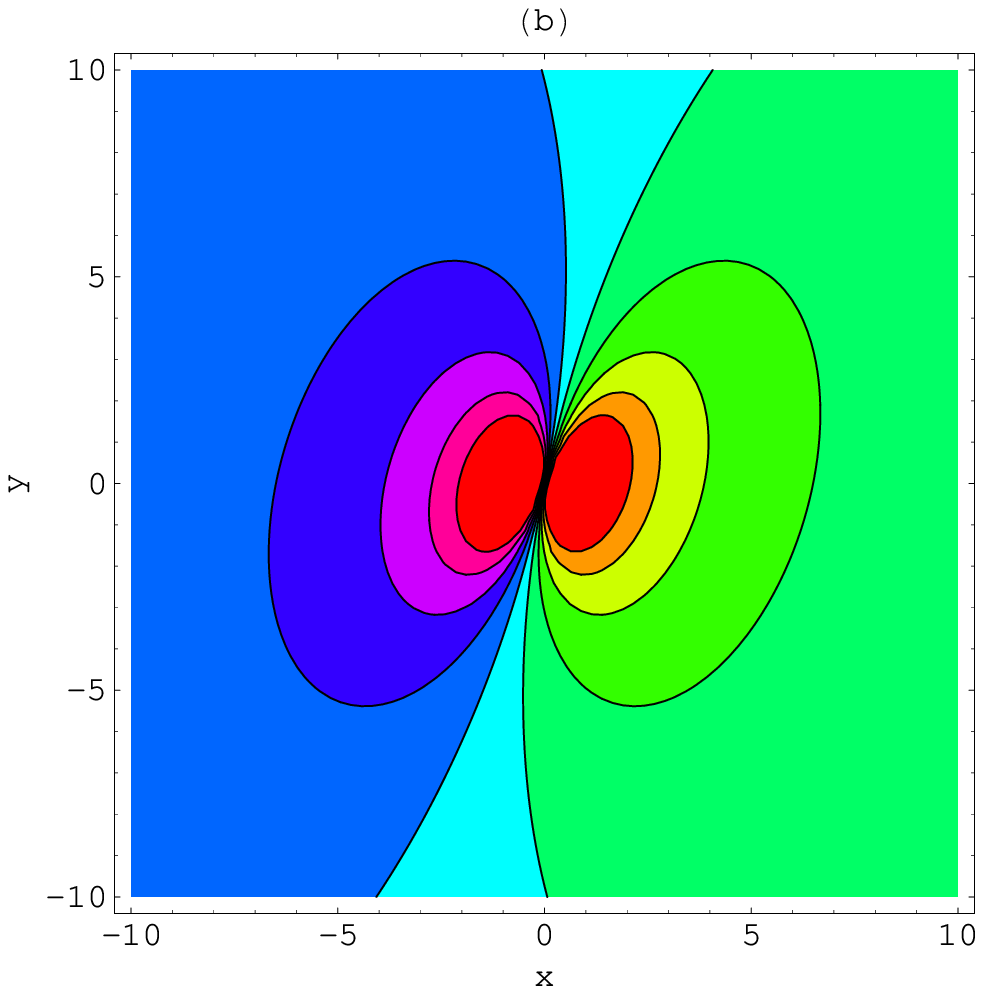}
\vspace{5cm}
\begin{tabbing}
\textbf{Fig. 1}. Lump solution (13) with $\alpha_1=\beta_3=-2$,
$\beta_4=-3$, $\alpha_4=2$,  $t=z=0$,\\
$\alpha_2=\beta_1=\gamma_3=1$.
\end{tabbing}

\section{Interaction
solutions between lump wave and solitary waves} \label{sec:3} \quad
In this section, we will discuss the interaction phenomenon between
lump wave and solitary waves. Considering the following mixed
functions {\begin{eqnarray} \xi&=&\gamma _3+\left(\alpha _4 t+\alpha
_1 x+\alpha _2 y+\alpha _3 z\right){}^2+\left(\beta _4 t+\beta
   _1 x+\beta _2 y+\beta _3 z\right){}^2\nonumber\\&+&\gamma _1 e^{t \varphi _4+\varphi _1 x+\varphi _2
   y+\varphi _3 z}+\gamma _2 e^{-t \varphi _4-\varphi _1 x-\varphi _2 y-\varphi _3 z},
\end{eqnarray}}where $\varphi_i(1\leq i \leq 4)$ and $\gamma_i(i=1,2)$ are
undetermined real parameters. As an example, substituting  Eq. (14)
and Eq. (5) into Eq. (3) and making the coefficients of $e^{t
\varphi _4+\varphi _1 x+\varphi _2
   y+\varphi _3 z} x^2$,
$e^{t \varphi _4+\varphi _1 x+\varphi _2
   y+\varphi _3 z} y^2$ et al. be zero, undetermined real parameters in
Eq. (14) have the following results
\begin{eqnarray} \alpha_3=-\alpha _1-\alpha _2, \beta_2= -\beta _1-\beta
_3, \varphi_3=-\varphi _1-\varphi _2.
\end{eqnarray}Substituting  Eq. (15) into Eq. (2) and Eq. (14), the interaction
solutions between lump wave and solitary waves can be read as
\begin{eqnarray}u&=&-[2 [-\gamma _2 \varphi _1 \exp [-t \varphi _4-\varphi _1 x-\varphi _2
   y+\left(\varphi _1+\varphi _2\right) z]+2 \alpha _1 [\alpha _4 t+\alpha _1 x\nonumber\\&+&\alpha
   _2 y-\left(\alpha _1+\alpha _2\right) z]+2 \beta _1 [\beta _4 t+\beta _1
   x-\left(\beta _1+\beta _3\right) y+\beta _3 z]\nonumber\\&+&\gamma _1 \varphi _1 e^{t \varphi
   _4+\varphi _1 x+\varphi _2 y-\left(\varphi _1+\varphi _2\right) z}]]/[\gamma _3+\gamma _2
   \exp [-t \varphi _4-\varphi _1 x-\varphi _2 y\nonumber\\&+&\left(\varphi _1+\varphi _2\right)
   z]+[\alpha _4 t+\alpha _1 x+\alpha _2 y-\left(\alpha _1+\alpha _2\right)
   z]{}^2+[\beta _4 t+\beta _1 x\nonumber\\&-&\left(\beta _1+\beta _3\right) y+\beta _3
   z]{}^2+\gamma _1 e^{t \varphi _4+\varphi _1 x+\varphi _2 y-\left(\varphi _1+\varphi
   _2\right) z}].
\end{eqnarray}Graphical representation of the interaction
solutions (16) is shown in Fig. 2, Fig. 3 and Fig. 4. Fig. 2
describes the interaction phenomenon between lump wave and one
solitary wave. Fig. 3 and Fig. 4 represent the interaction
phenomenon between lump wave and two solitary waves.

\includegraphics[scale=0.4,bb=20 270 10 10]{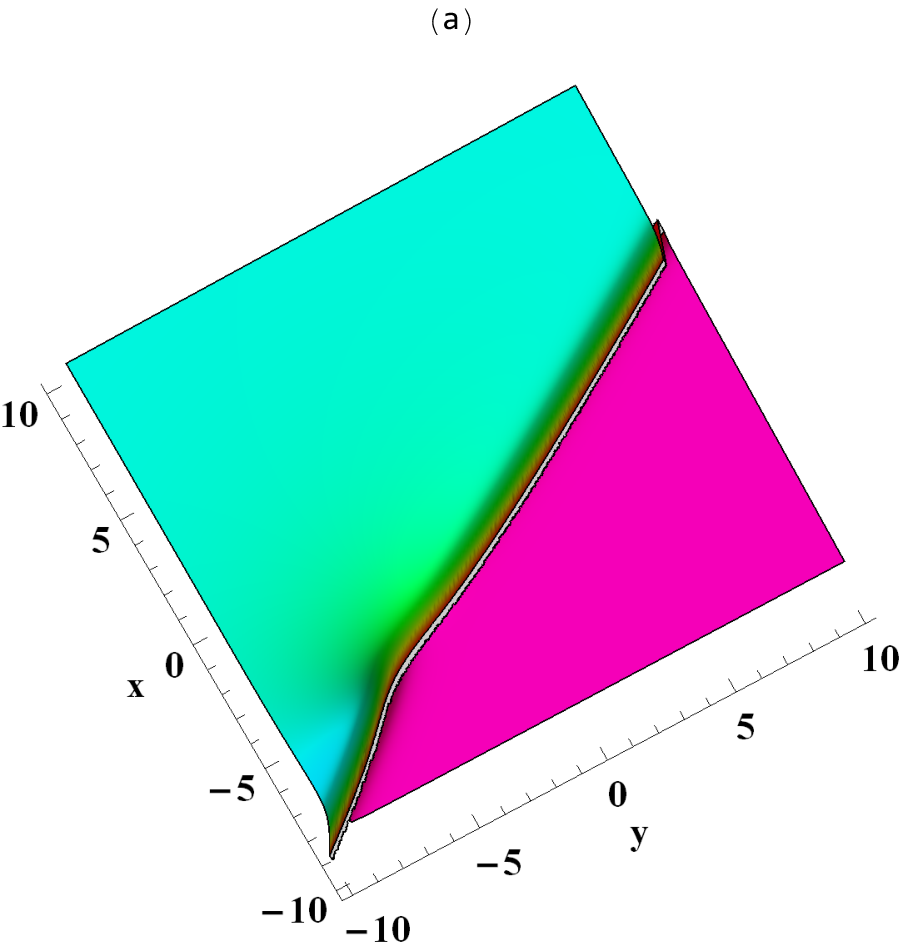}
\includegraphics[scale=0.4,bb=-255 270 10 10]{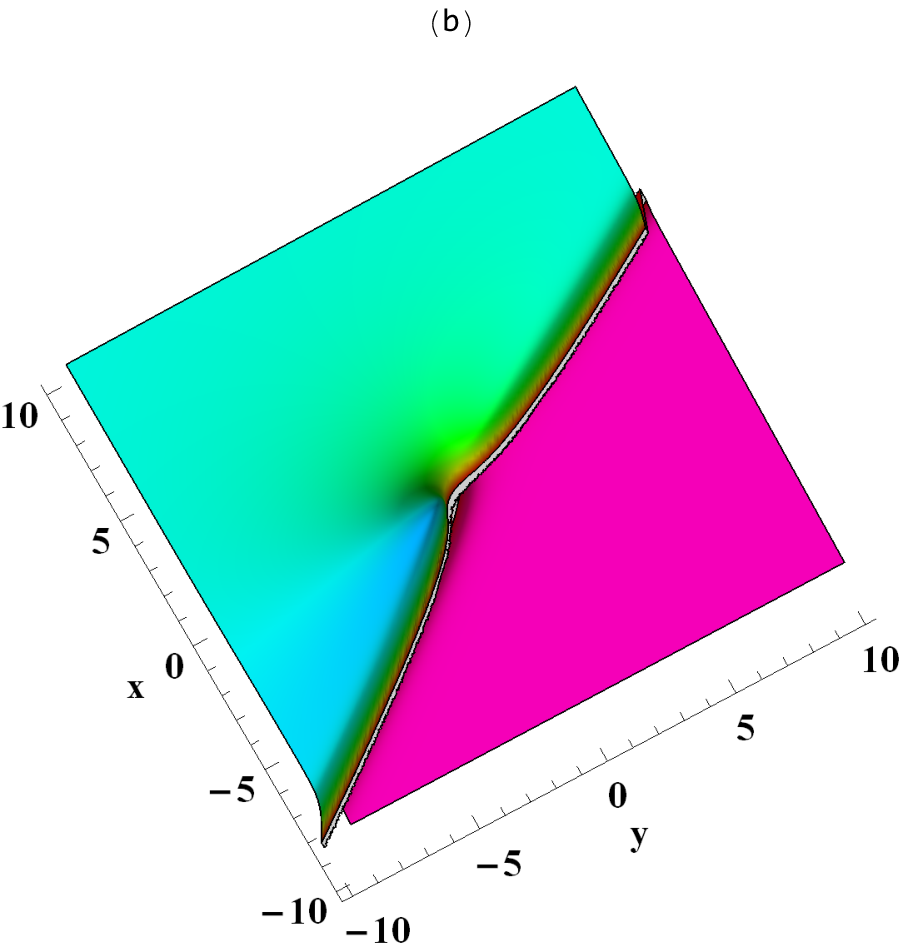}
\includegraphics[scale=0.4,bb=-260 270 10 10]{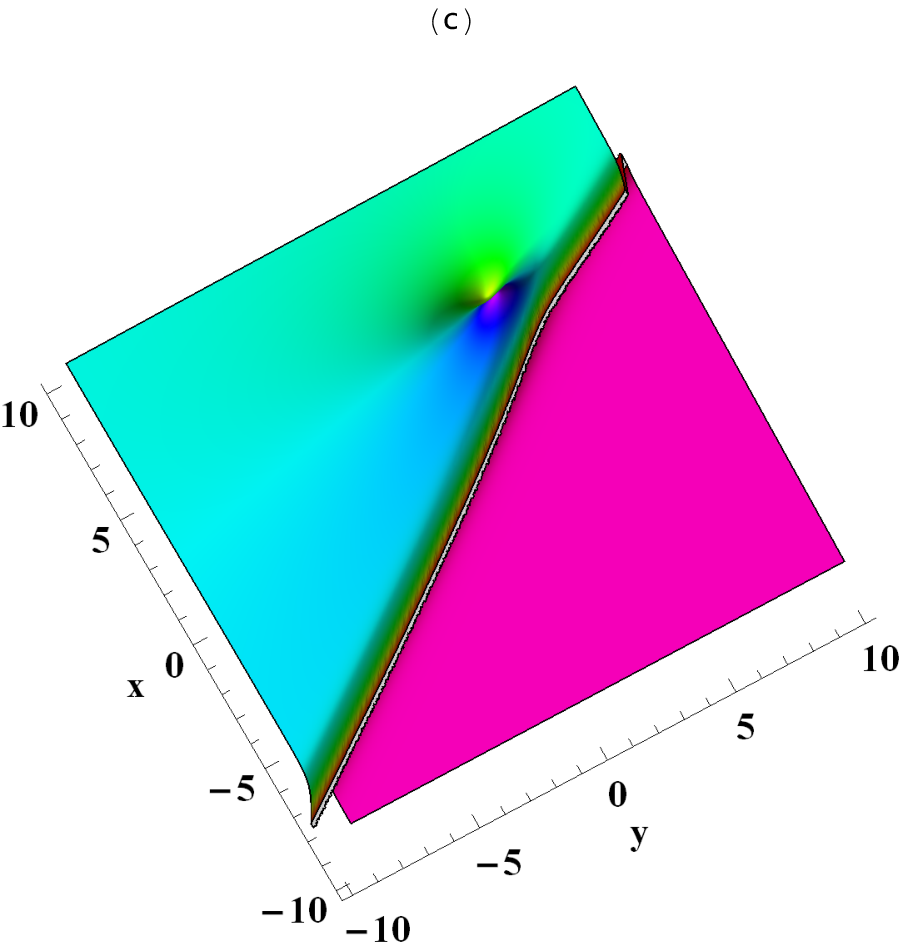}
\includegraphics[scale=0.35,bb=750 620 10 10]{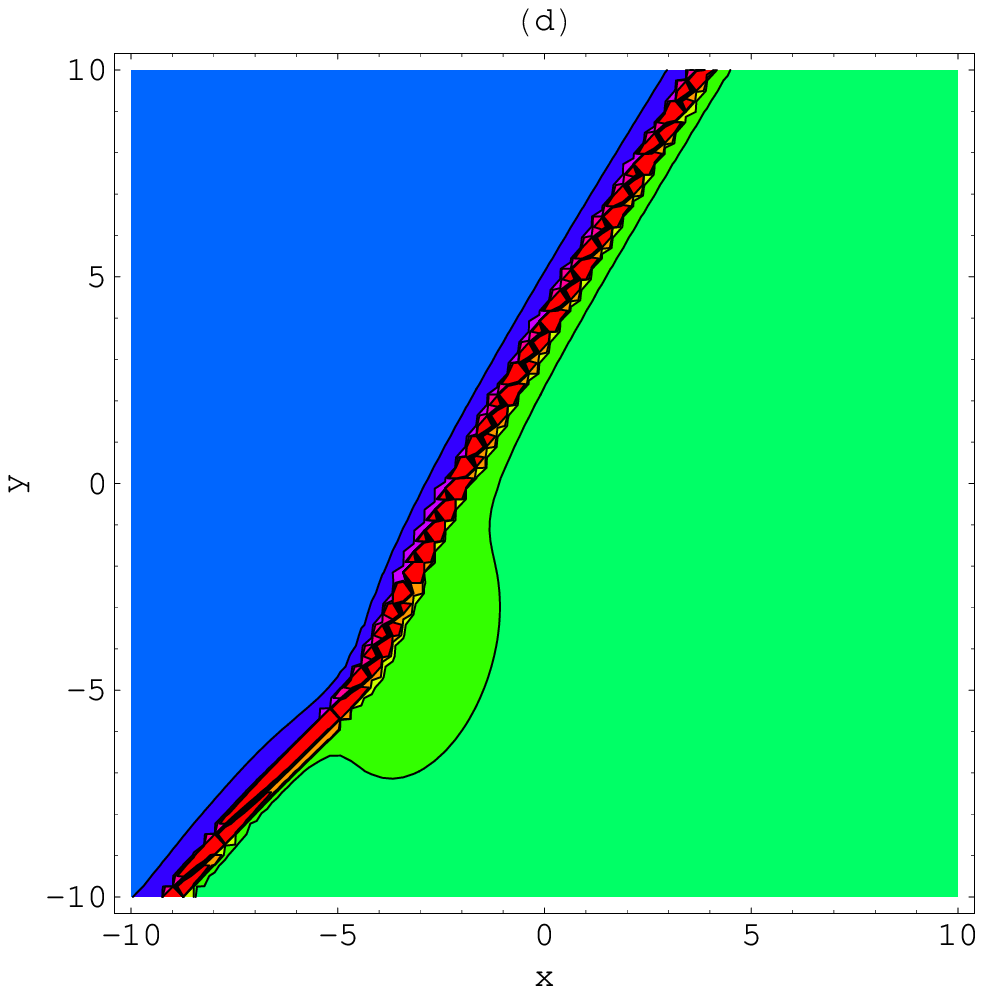}
\includegraphics[scale=0.35,bb=-290 620 10 10]{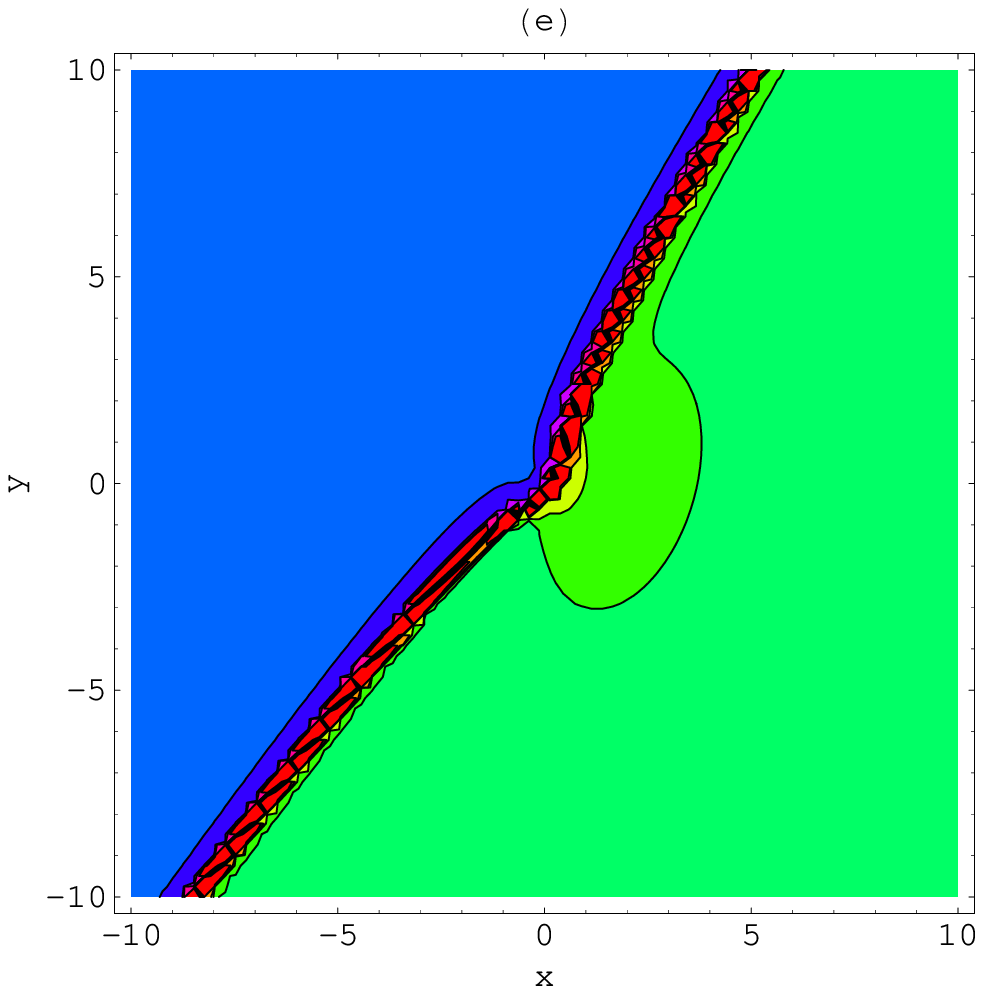}
\includegraphics[scale=0.35,bb=-290 620 10 10]{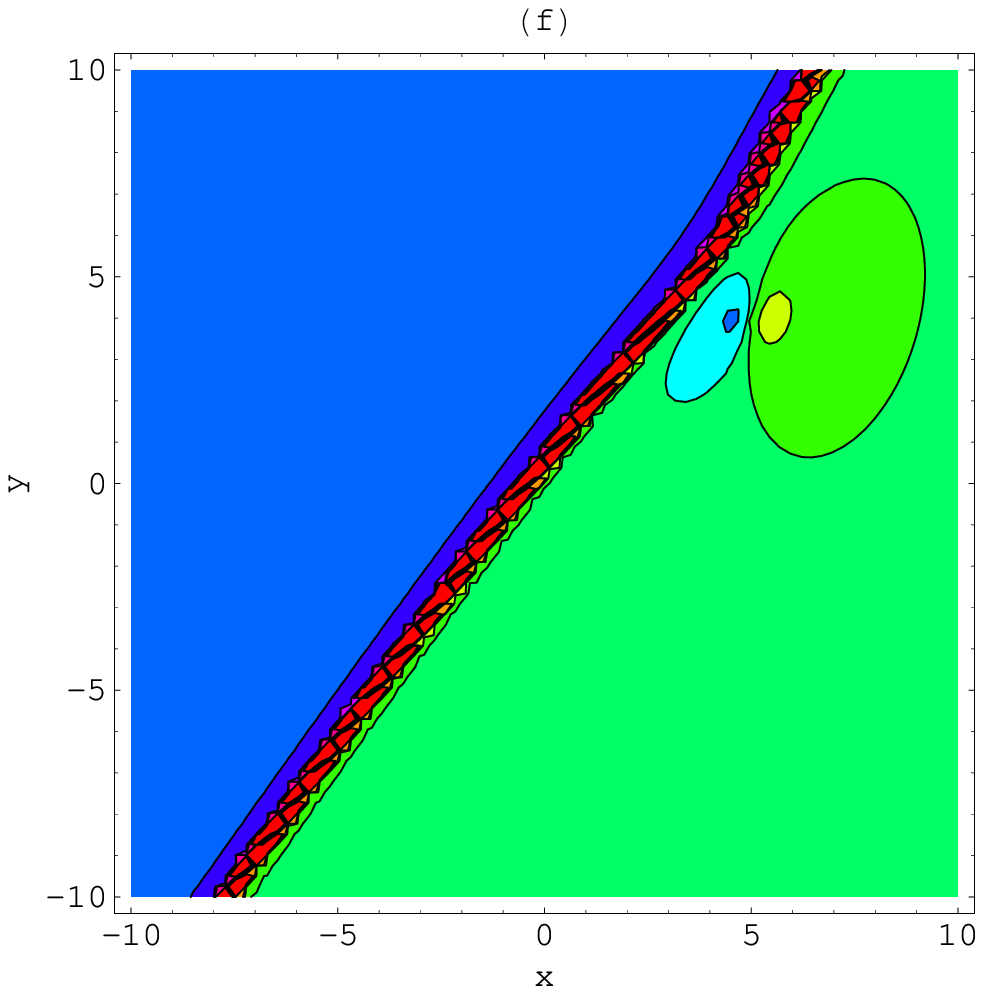}
\vspace{7.4cm}
\begin{tabbing}
\textbf{Fig. 2}.  Solution (16) with $\alpha_1=\beta_3=\gamma_1=-2$,
$\beta_4=\varphi_1=-3$, $\alpha_4=\varphi_2=2$,\\  $\gamma_2=z=0$,
$\alpha_2=\beta_1=\gamma_3=\varphi_4=1$, when $t=-3$ in (a,d), $t=0$
in (b,e),\\ $t=3$ in (c,f).
\end{tabbing}

\section{Interaction
solutions between lump wave and periodic waves} \label{sec:4} \quad
In this section, we will investigate the interaction phenomenon
between lump wave and periodic waves. Assuming the following mixed
functions {\begin{eqnarray} \xi&=&\gamma _3+\left(\alpha _4 t+\alpha
_1 x+\alpha _2 y+\alpha _3 z\right){}^2+\left(\beta _4 t+\beta
   _1 x+\beta _2 y+\beta _3 z\right){}^2\nonumber\\&+&\gamma _2 \sin \left(\mu _4 t+\mu _1 x+\mu _2 y+\mu _3
   z\right)+\gamma _1 \cos \left(t \varphi _4+\varphi _1 x+\varphi _2 y+\varphi _3 z\right),
\end{eqnarray}}where $\mu_i(1\leq i \leq 4)$ is
undetermined real parameter.

\includegraphics[scale=0.4,bb=20 260 10 10]{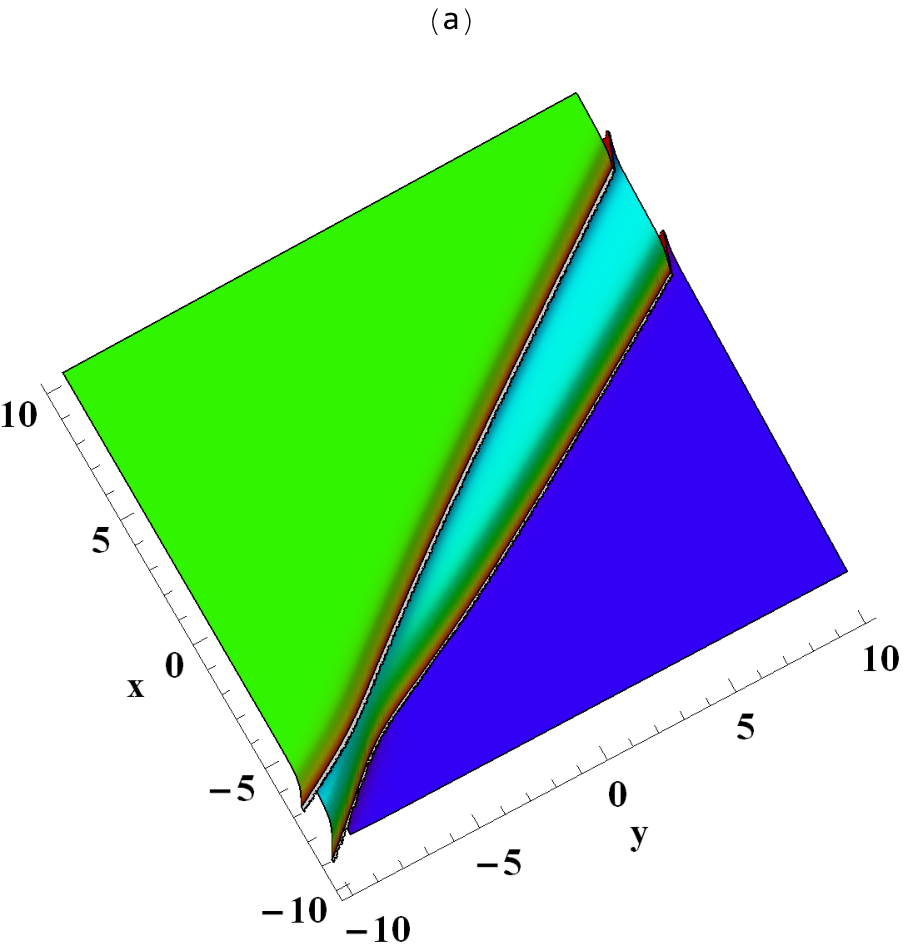}
\includegraphics[scale=0.4,bb=-255 260 10 10]{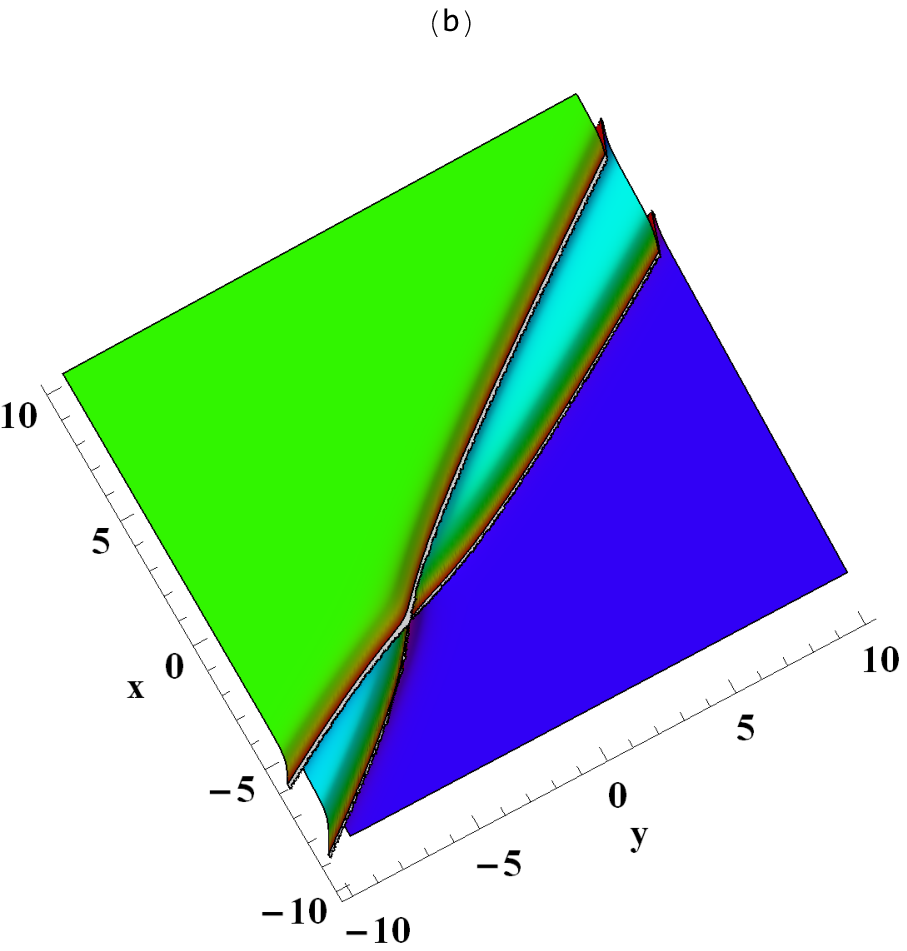}
\includegraphics[scale=0.4,bb=-260 260 10 10]{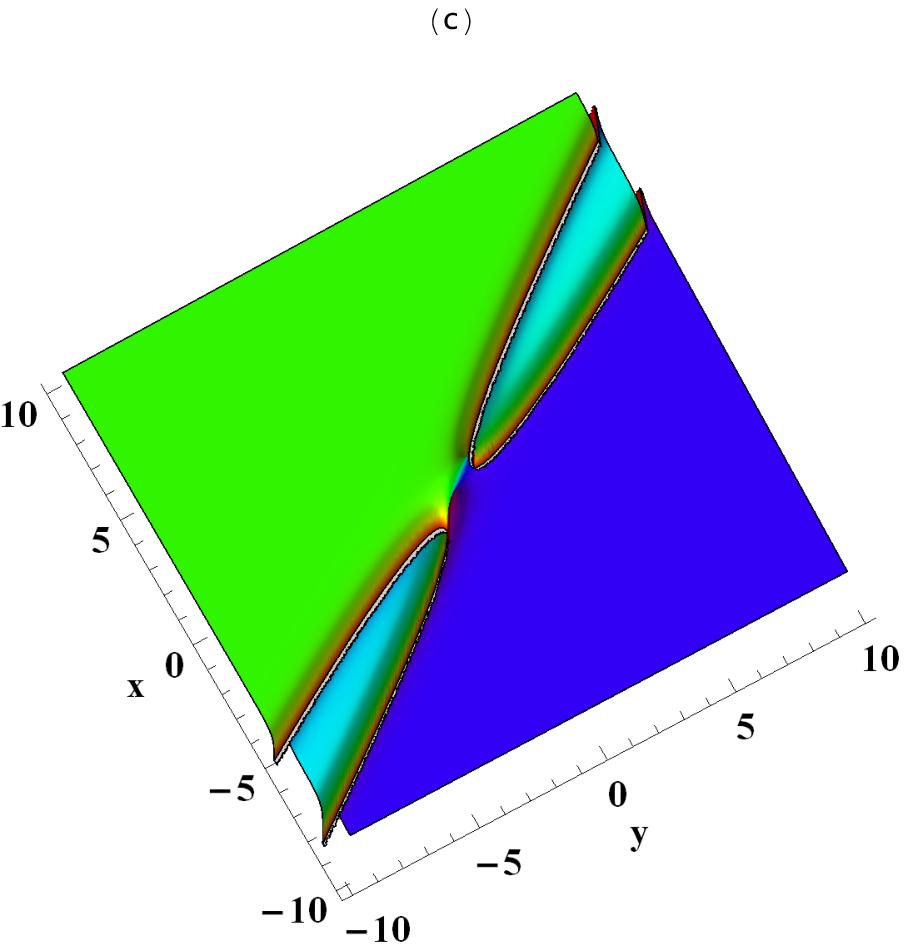}
\includegraphics[scale=0.4,bb=450 525 10 10]{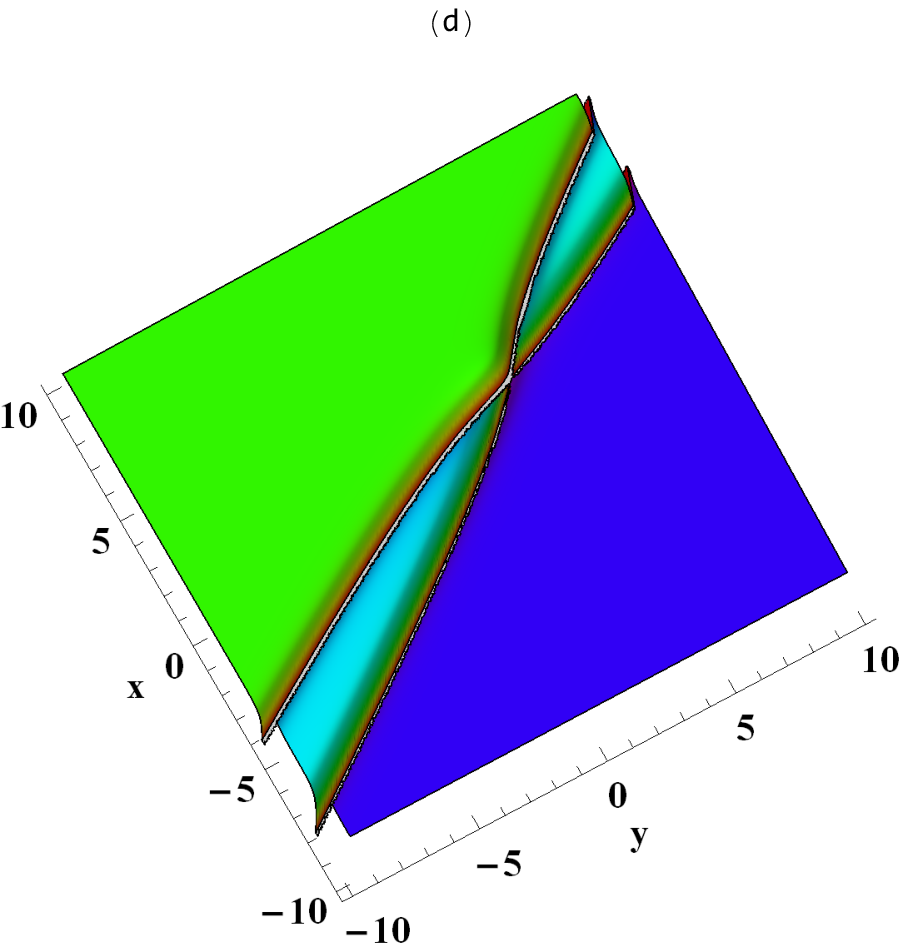}
\includegraphics[scale=0.4,bb=-300 525 10 10]{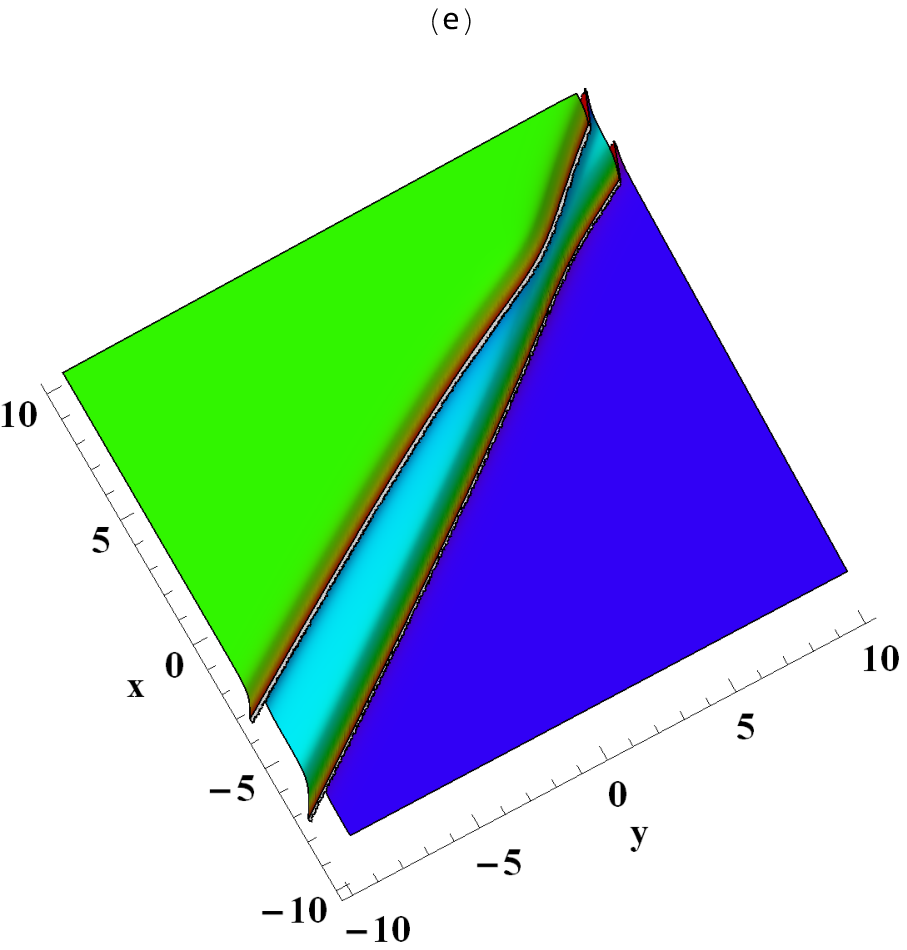}
\vspace{7.4cm}
\begin{tabbing}
\textbf{Fig. 3}. Solution (16) with
$\alpha_1=\beta_3=\gamma_1=\gamma_2=-2$, $\beta_4=\varphi_1=-3$,
$z=0$,\\ $\alpha_4=\varphi_2=2$,
$\alpha_2=\beta_1=\gamma_3=\varphi_4=1$, when $t=-4$ in (a), $t =
-2$ in (b),\\ $t = 0$ in (c), $t = 2$ in (d), $t = 4$ in (e).
\end{tabbing}

\includegraphics[scale=0.35,bb=140 290 10 10]{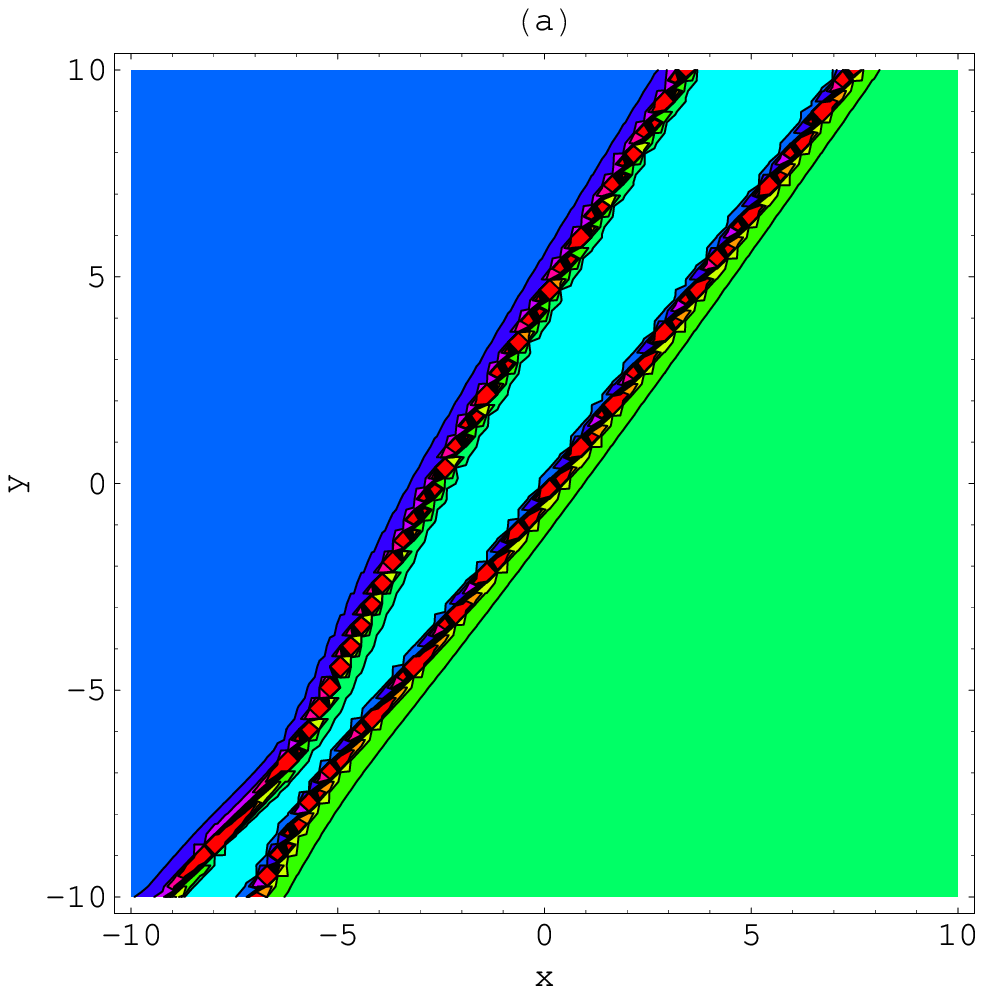}
\includegraphics[scale=0.35,bb=-275 290 10 10]{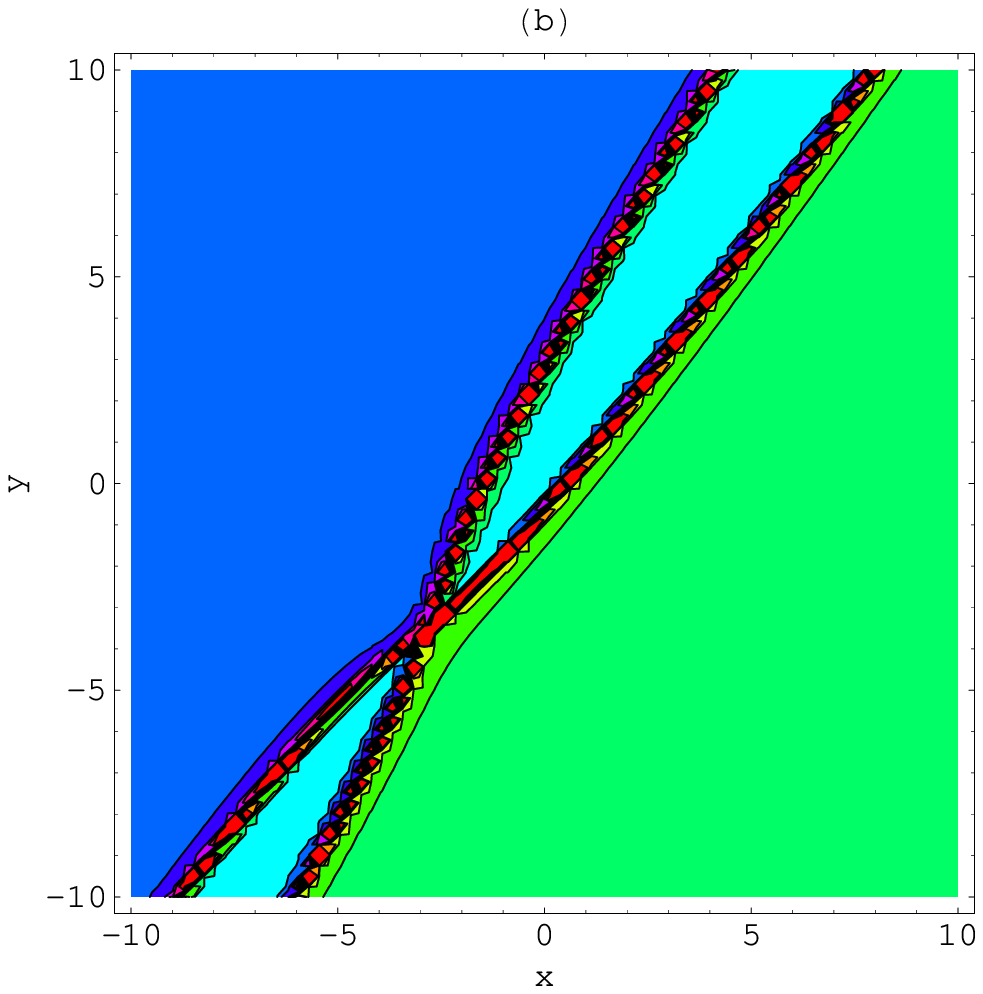}
\includegraphics[scale=0.35,bb=-280 290 10 10]{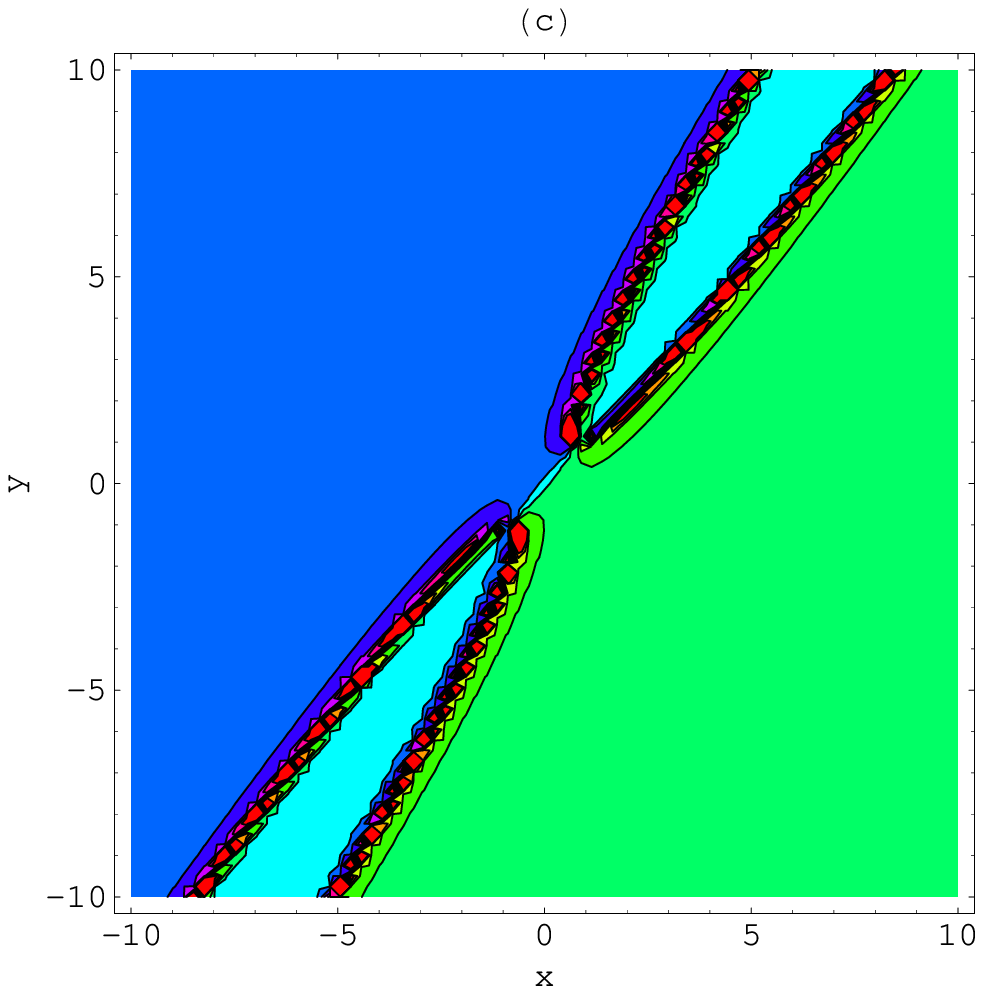}
\includegraphics[scale=0.35,bb=500 585 10 10]{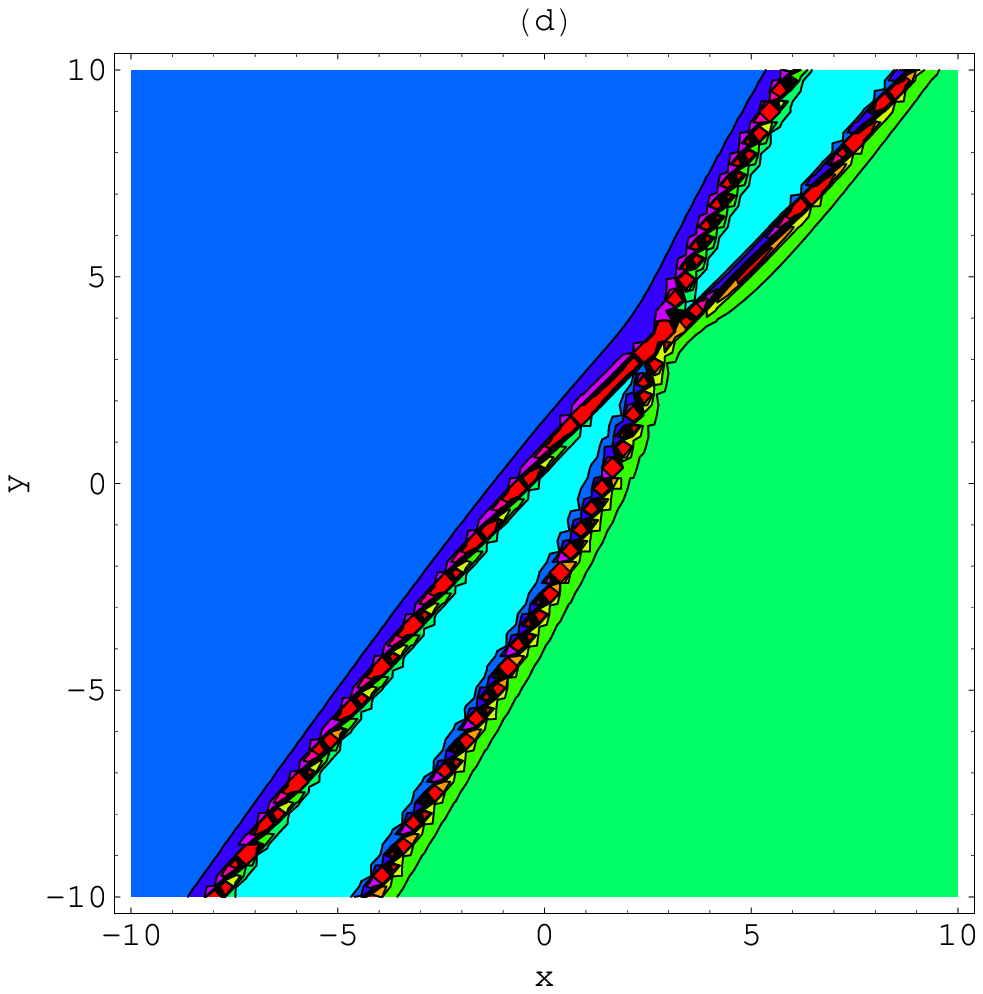}
\includegraphics[scale=0.35,bb=-360 585 10 10]{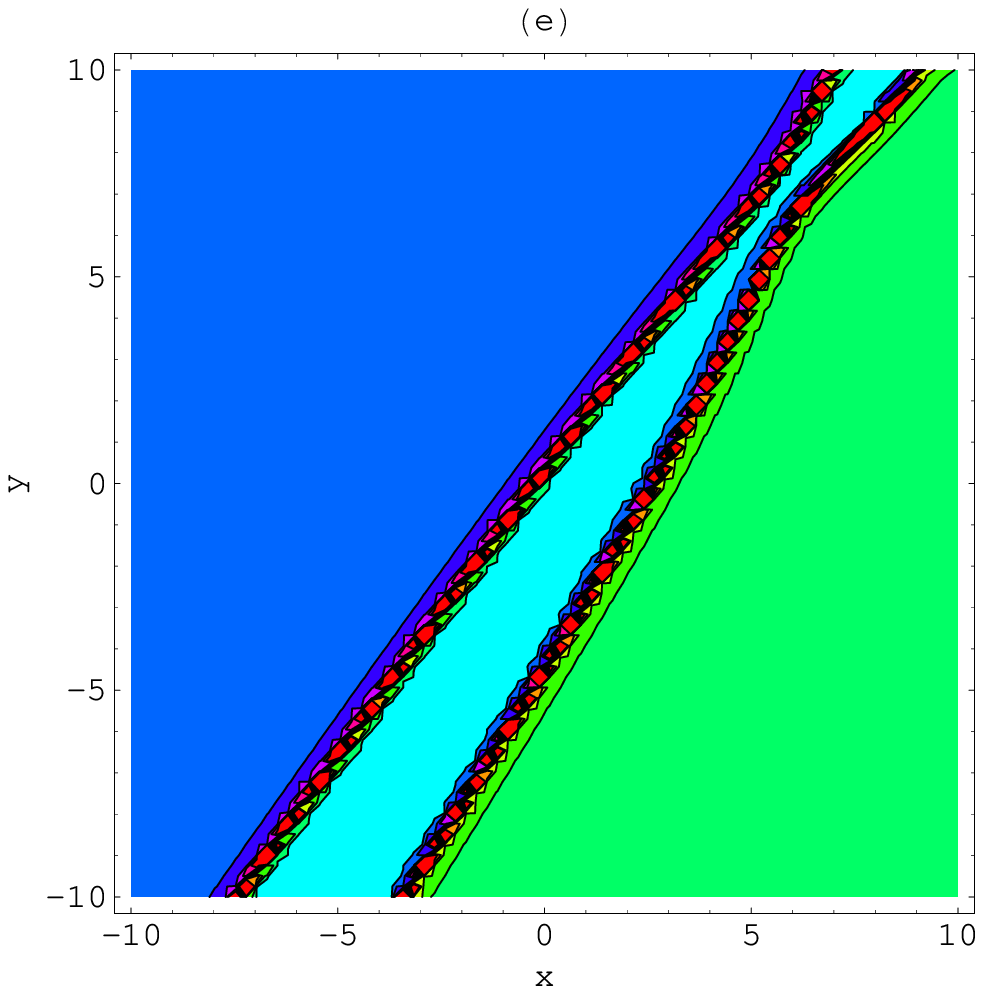}
\vspace{6.cm}
\begin{tabbing}
\textbf{Fig. 4}. The corresponding contour plots of Fig. 3.\\
\end{tabbing}
Substituting  Eq. (17) and Eq. (5) into Eq. (3) and making the
coefficients of $\cos \left(t \varphi _4+\varphi _1 x+\varphi _2
y+\varphi _3 z\right) x^2$, $\sin \left(\mu _4 t+\mu _1 x+\mu _2
y+\mu _3
   z\right) x^2$ et al. be zero, undetermined real parameters in
Eq. (17) have the following results
\begin{eqnarray} \alpha_3=-\alpha _1-\alpha _2, \beta_2= -\beta _1-\beta
_3, \varphi_3=-\varphi _1-\varphi _2, \mu_3=-\mu _1-\mu _2.
\end{eqnarray}Substituting  Eq. (18) into Eq. (2) and Eq. (17), the interaction
solutions between lump wave and periodic waves can be written as
\begin{eqnarray}u&=&-[2 [2 \alpha _1 [\alpha _4 t+\alpha _1 x+\alpha _2 y-\left(\alpha _1+\alpha
   _2\right) z]+2 \beta _1 [\beta _4 t+\beta _1 x-\left(\beta _1+\beta _3\right) y\nonumber\\&+&\beta
   _3 z]+\gamma _2 \mu _1 \cos [\mu _4 t+\mu _1 x+\mu _2 y-\left(\mu _1+\mu _2\right)
   z]-\gamma _1 \varphi _1 \sin [t \varphi _4+\varphi _1 x\nonumber\\&+&\varphi _2 y-\left(\varphi
   _1+\varphi _2\right) z]]]/[\gamma _3+[\alpha _4 t+\alpha _1 x+\alpha _2
   y-\left(\alpha _1+\alpha _2\right) z]{}^2+[\beta _4 t\nonumber\\&+&\beta _1 x-\left(\beta _1+\beta
   _3\right) y+\beta _3 z]{}^2+\gamma _2 \sin [\mu _4 t+\mu _1 x+\mu _2 y-\left(\mu
   _1+\mu _2\right) z]\nonumber\\&+&\gamma _1 \cos [t \varphi _4+\varphi _1 x+\varphi _2
   y-\left(\varphi _1+\varphi _2\right) z]].
\end{eqnarray}

\includegraphics[scale=0.55,bb=-20 270 10 10]{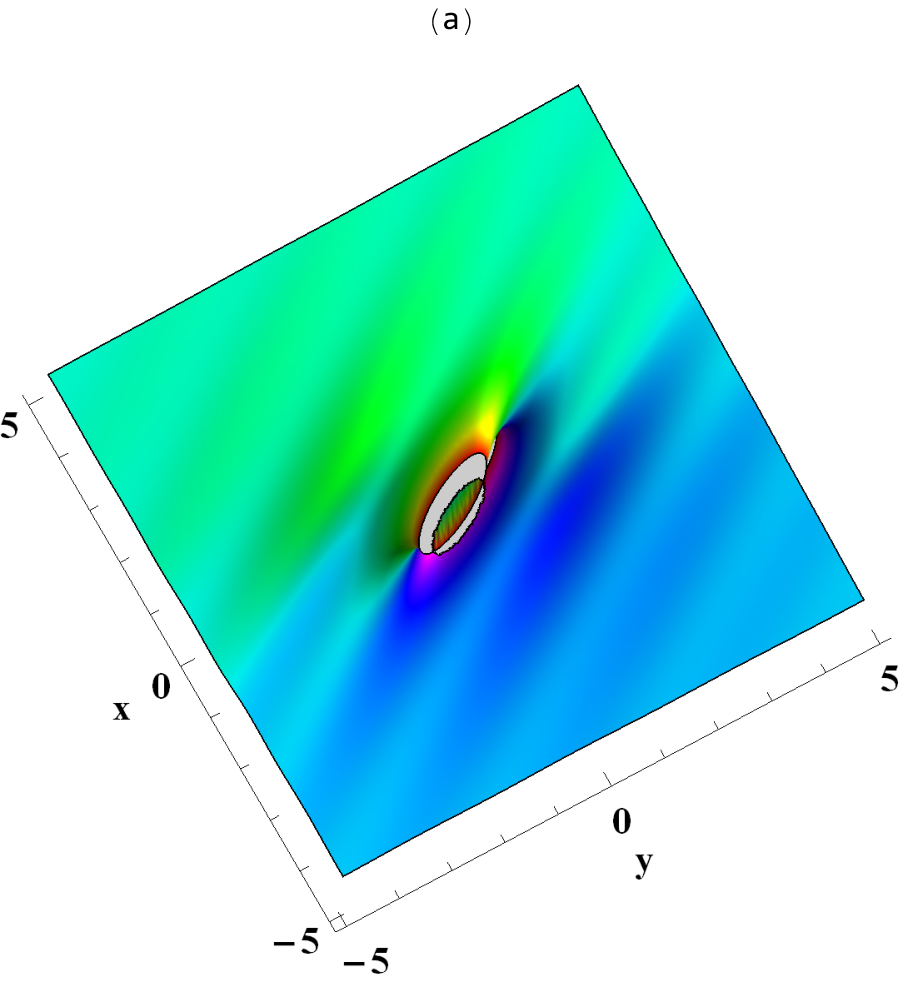}
\includegraphics[scale=0.45,bb=-270 310 10 10]{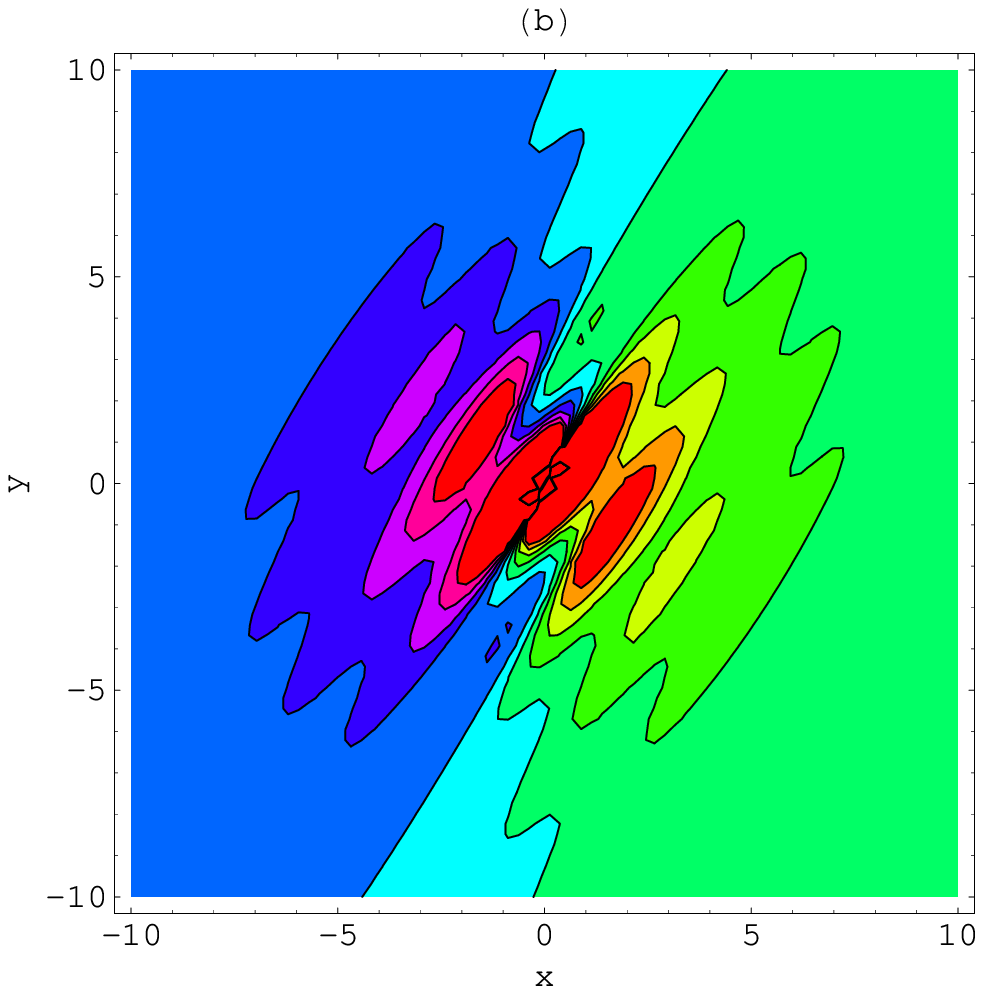}
\vspace{5.2cm}
\begin{tabbing}
\textbf{Fig. 5}. Solution (19) with $\alpha_1=\beta_3=\gamma_1=-2$,
$\beta_4=\varphi_1=-3$, $z=t=\gamma_2=0$,\\
$\alpha_4=\varphi_2=\mu_1=\mu_4=2$,
$\alpha_2=\beta_1=\gamma_3=\varphi_4=1$, $\mu_2=-1$.
\end{tabbing}

\includegraphics[scale=0.55,bb=-20 270 10 10]{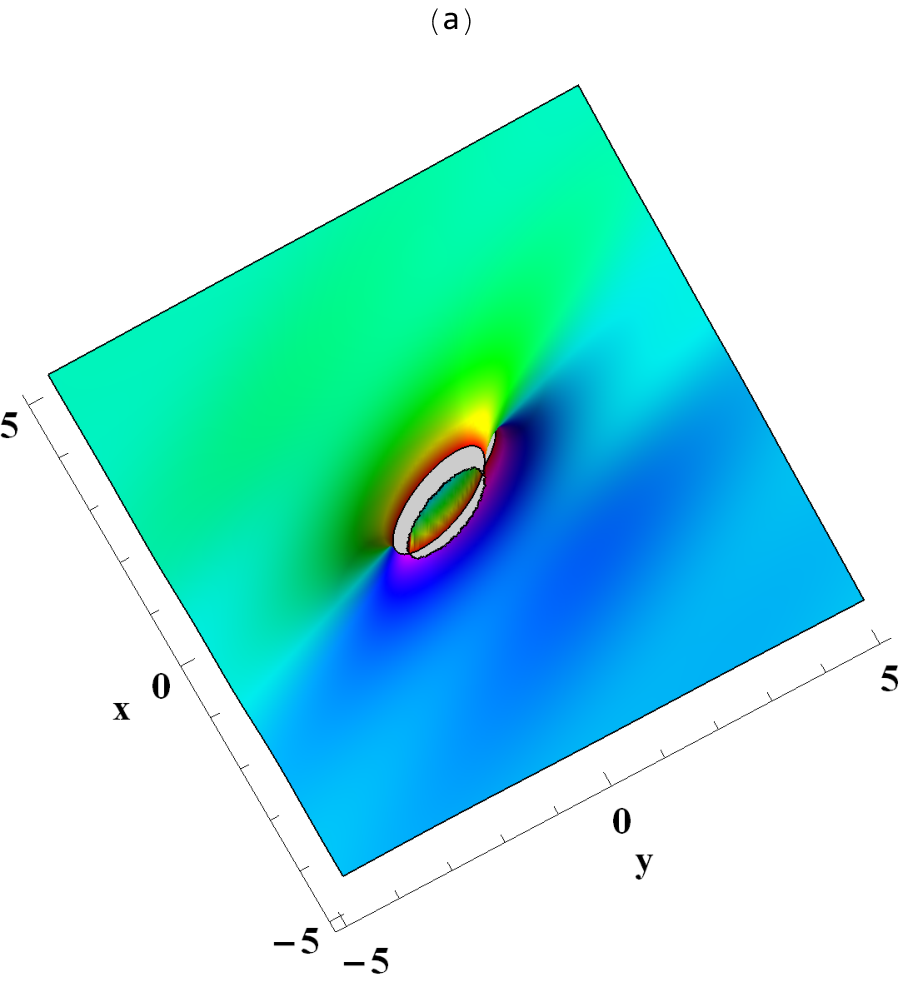}
\includegraphics[scale=0.45,bb=-270 310 10 10]{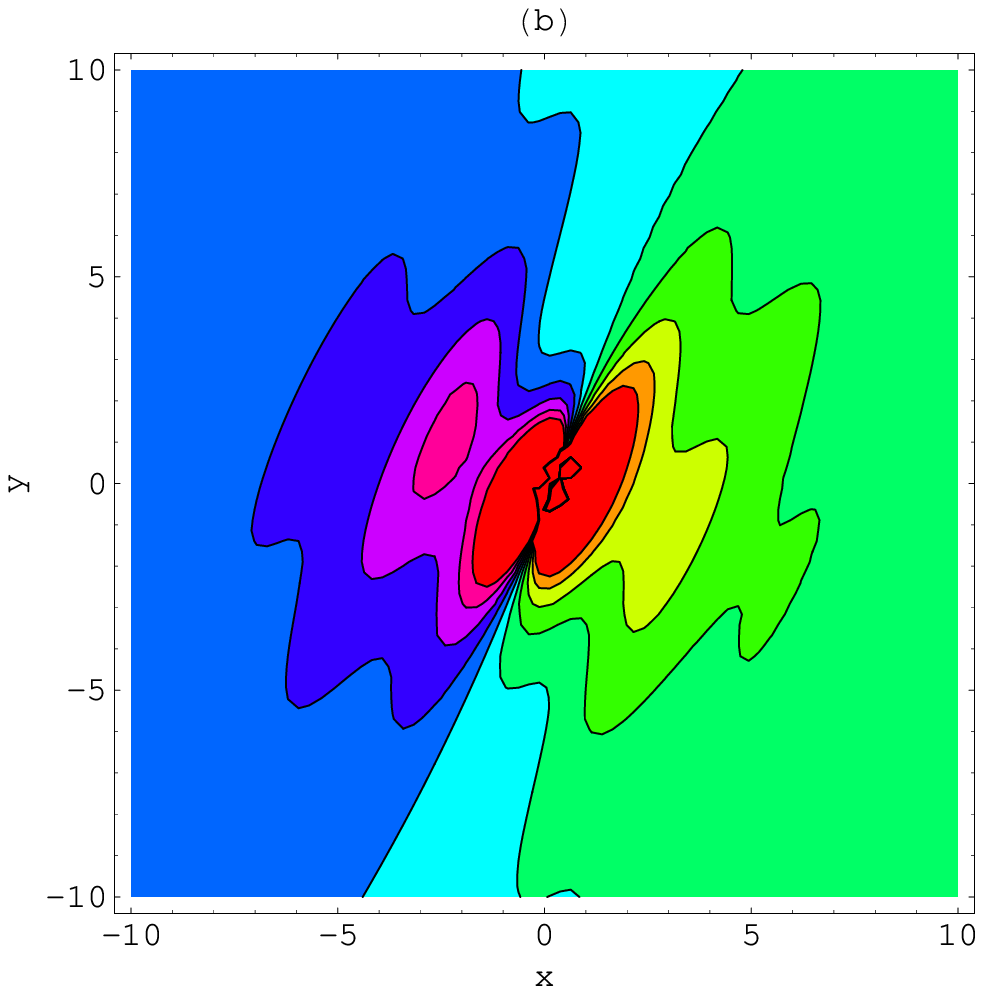}
\vspace{5.2cm}
\begin{tabbing}
\textbf{Fig. 6}. Solution (19) with
$\alpha_1=\beta_3=\gamma_1=\gamma_2=-2$,
$\beta_4=-3$, $z=t=0$,\\
$\alpha_4=\mu_1=\mu_4=\varphi_1=\varphi_4=2$,
$\alpha_2=\beta_1=\gamma_3=1$, $\mu_2=\varphi_2=-1$.\\
\end{tabbing}Graphical representation of the interaction
solutions (19) is shown in Fig. 5 and Fig. 6. Fig. 5 describes the
interaction phenomenon between lump wave and one periodic wave. Fig.
6 represents the interaction phenomenon between lump wave and two
periodic waves.

\section{Breather wave solutions} \label{sec:5}
\quad In this section, we will study the breather wave solutions.
Choosing the following mixed functions{\begin{eqnarray} \xi&=&k_1
e^{t \varphi _4+\varphi _1 x+\varphi _2 y+\varphi _3 z}+\gamma _2
\sin \left(\mu _4 t+\mu _1
   x+\mu _2 y+\mu _3 z\right)\nonumber\\&+&\gamma _1 \cos \left(\nu _4 t+\nu _1 x+\nu _2 y+\nu _3 z\right)+e^{-t
   \varphi _4-\varphi _1 x-\varphi _2 y-\varphi _3 z},
\end{eqnarray}}where  $\nu_i(1\leq i \leq 6)$ and $k_1$ are
unknown real parameters. Substituting  Eq. (20) into Eq. (3), we
have the following results
\begin{eqnarray} \nu_3=-\nu _1-\nu _2, \varphi_3=-\varphi _1-\varphi _2, \mu_3=-\mu _1-\mu _2.
\end{eqnarray}Substituting  Eq. (21) into Eq. (2) and Eq. (20), the breather wave solutions can be presented as
\begin{eqnarray}u&=&-[2 [-\varphi _1 \exp [-t \varphi _4-\varphi _1 x-\varphi _2 y+\left(\varphi
   _1+\varphi _2\right) z]+\gamma _2 \mu _1 \cos [\mu _4 t+\mu _1 x\nonumber\\&+&\mu _2
   y-\left(\mu _1+\mu _2\right) z]+k_1 \varphi _1 e^{t \varphi _4+\varphi _1 x+\varphi _2
   y-\left(\varphi _1+\varphi _2\right) z}-\gamma _1 \nu _1 \sin [\nu _4 t+\nu _1 x\nonumber\\&+&\nu _2
   y-\left(\nu _1+\nu _2\right) z]]]/[\exp [-t \varphi _4-\varphi _1 x-\varphi
   _2 y+\left(\varphi _1+\varphi _2\right) z]\nonumber\\&+&k_1 e^{t \varphi _4+\varphi _1 x+\varphi _2
   y-\left(\varphi _1+\varphi _2\right) z}+\gamma _2 \sin [\mu _4 t+\mu _1 x+\mu _2
   y-\left(\mu _1+\mu _2\right) z]\nonumber\\&+&\gamma _1 \cos [\nu _4 t+\nu _1 x+\nu _2 y-\left(\nu
   _1+\nu _2\right) z]].
\end{eqnarray}Graphical representation of the breather wave solutions (22) is displayed in Fig.
7.

\includegraphics[scale=0.55,bb=-20 270 10 10]{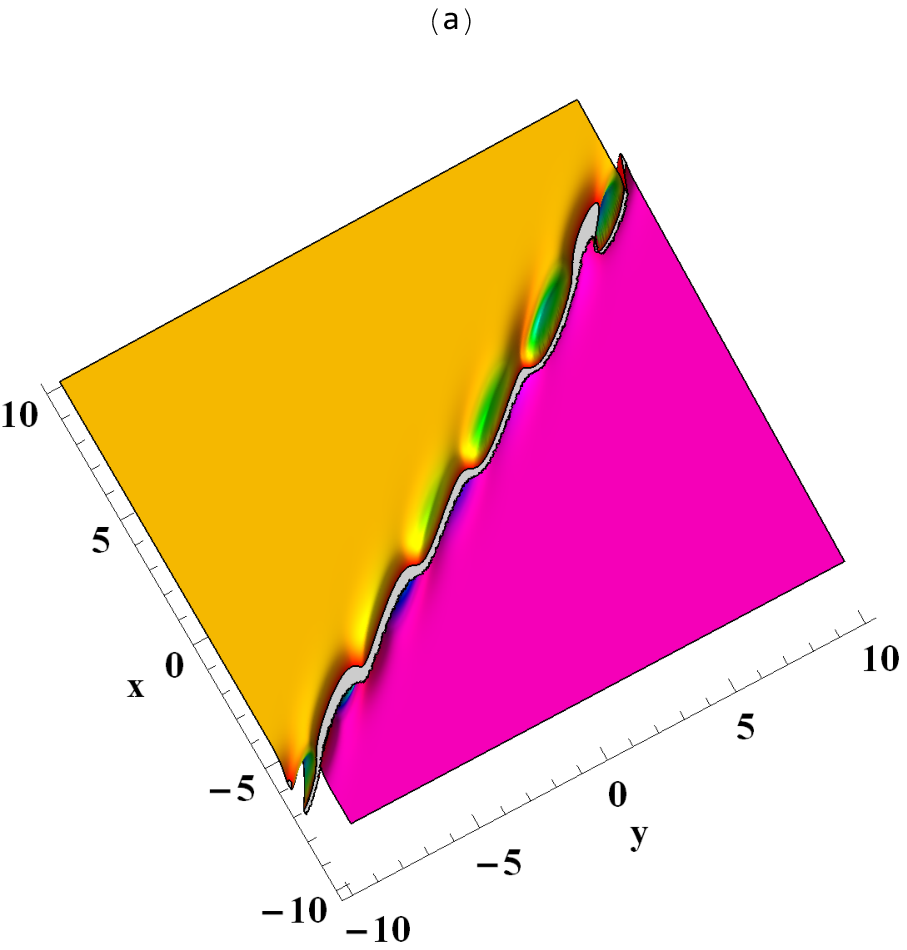}
\includegraphics[scale=0.45,bb=-270 310 10 10]{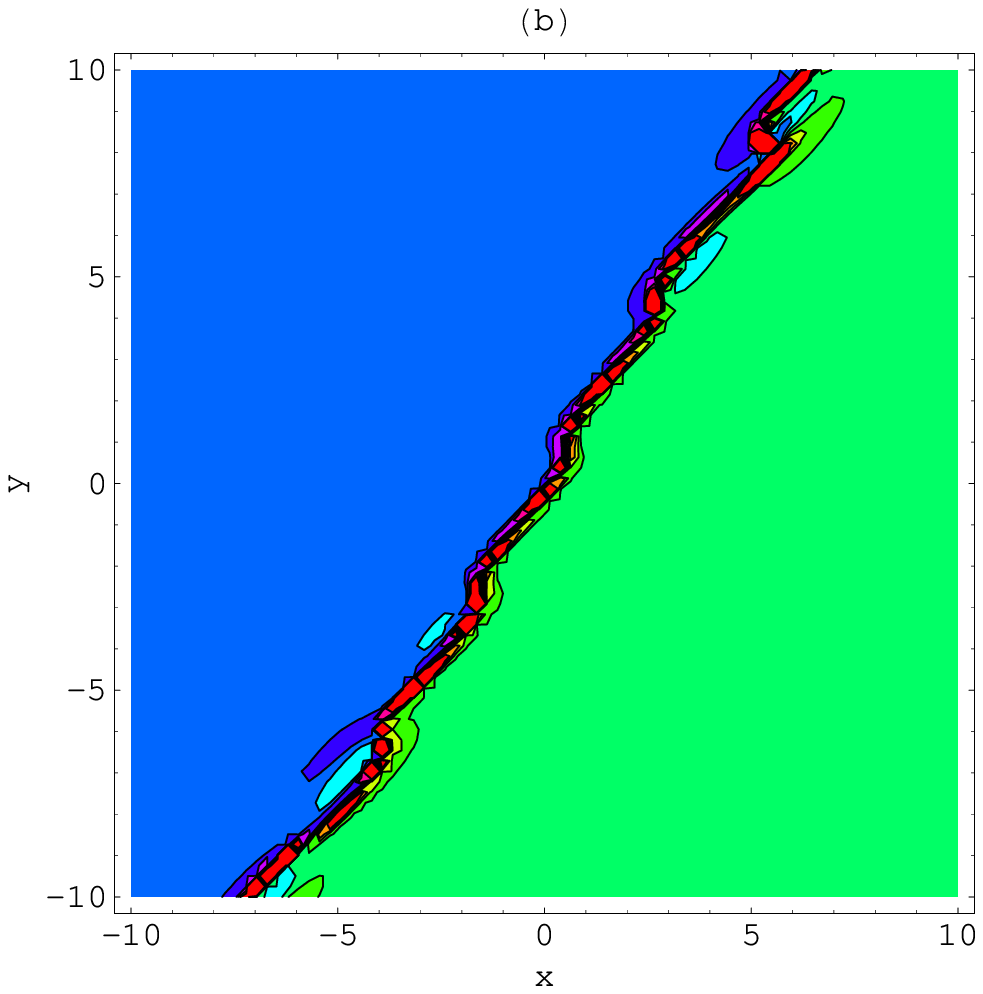}
\vspace{5.2cm}
\begin{tabbing}
\textbf{Fig. 7}.  Breather wave solutions (22) with $\gamma_1=-2$,
$\varphi_1=\nu_2=-3$, $z=t=0$,\\
$\gamma_2=\mu_1=\mu_4=\varphi_2=\nu_1=2$, $\nu_4=\varphi_4=1$,
$\mu_2=k_1=-1$.
\end{tabbing}

\section{ Conclusion}
\label{sec:3} \quad Recently, a new (3+1)-dimensional BLMP equation
is introduced by Wazwaz. At present, there is no literature on the
lump solution of this equation. In this paper, breather wave and
lump-type solutions of the new (3+1)-dimensional BLMP equation
 are presented. Lump-type solutions contain the interaction solutions between lump and solitary waves,
 and the interaction solutions between lump and periodic waves, which have not been studied in any literature.
 Furthermore, graphical representation for lump solution is displayed in Fig. 1. Interaction phenomenon of
 lump-type
 solutions are demonstrated in Figs. 2-7.\\

\noindent {\bf Compliance with ethical standards}\\

\quad {\bf Conflict of interests} The authors declare that there is
no conflict of interests regarding the publication of this article.

\quad {\bf Ethical standard} The authors state that this research
complies with ethical standards. This research does not involve
either human participants or animals.



\end{document}